# Spin Polarization Inversion at Benzene-Absorbed $Fe_4N$ Surface

Qian Zhang[1], Wenbo Mi[1,*], Xiaocha Wang[2], and Xuhui Wang[3,*]

*[1]Tianjin Key Laboratory of Low Dimensional Materials Physics and Preparation Technology, Faculty of Science, Tianjin University, Tianjin 300072, China*

*[2]Tianjin Key Laboratory of Film Electronic & Communicate Devices, School of Electronics Information Engineering, Tianjin University of Technology, Tianjin 300384, China*

*[3]Physical Science and Engineering Division, King Abdullah University of Science and Technology (KAUST), Thuwal 23955-6900, Kingdom of Saudi Arabia*

[*]Author to whom all correspondence should be addressed.

E-mail: miwenbo@tju.edu.cn and xuhuiwangnl@gmail.com

# ABSTRACT

We report a first-principle study on electronic structure and simulation of the spin-polarized scanning tunneling microscopy graphic of a benzene/$Fe_4N$ interface. $Fe_4N$ is a compound ferromagnet suitable for many spintronic applications. We found that, depending on the particular termination schemes and interface configurations, the spin polarization on the benzene surface shows a rich variety of properties ranging from cosine-type oscillation to polarization inversion. Spin-polarization inversion above benzene is resulting from the hybridizations between C $p_z$ and the out-of-plane $d$ orbitals of Fe atom.

**Keywords:** Spin-polarization inversion, Absorption, Interfaces

**PACS:** 72.25.-b, 68.43.Bc, 73.20.-r

# Introduction

In recent years, combining inorganic and organic materials have opened many venues for the novel science and applications. The rising field of organic spintronics is one of them[1-6]. Organic materials, with long spin-flip diffusion length and weak spin-orbit coupling, are desirable in many spintronic applications[2, 7-17]. Moreover, the functionalities of devices made of organic materials can be manipulated through relatively simple methods such as ligand modification[18, 19] and isomerization[20, 21].

A major topic in organic spintronics is the spin related properties at the interface between a ferromagnetic substrate and organic material, i.e. the spinterface[22]. Much effort has been made to clarify the underlying mechanisms that drive the peculiarities, such as spin polarization inversion, at various spinterfaces[21, 23-26]. For the interfaces between Fe and benzene molecule (as well as $C_5H_5$ and $C_8H_8$), Atodiresei *et al.* argue that it is the $p_z$-$d$ Zener exchange-type mechanism that leads to the spin-polarization inversion[24]. In another study on the thiophene/cobalt(001) interface[25], the strong spatial dependence of the spin polarization at the interface is attributed to the reduced molecular symmetry. The azobenzene isomer adsorbed on Fe surface has been reported recently, where the switch between two types of azobenzenes takes place by applying light and/or heat [21]. On the other hand, studies are extended to systems consisting of AFM substrates including benzene adsorbed on monolayer AFM Mn[26] or metal phthalocyanine[4, 27-29], where the spin polarization modification has often been reported.

But, we are not aware of any investigations into the interfaces between an organic molecule and a compound ferromagnet, such as iron nitride ($Fe_4N$). $Fe_4N$ carries a high spin polarization of nearly ~100% [30] as well as a large saturation magnetization of 1200 emu/cm$^3$ [31]. Its Curie temperature is about 760 K. Together with its high chemical stability[32] and low coercivity[33], $Fe_4N$ is a promising candidate for, among other, spin injection source[34]. In this work, we scrutinize the spinterface between a benzene molecule and $Fe_4N$ substrate. In particular, we will show that different termination schemes and adsorption configurations unique to the benzene/$Fe_4N$ interface enrich the properties of spin polarization.

## Calculation details

Our first principles calculations are based on the density-functional theory (DFT) and the projector augmented wave method as implemented in the Vienna Ab initio Simulation Package code[35, 36]. For the exchange and correlation functional, we use the Perdew-Burke-Ernzerhof spin-polarized generalized gradient approximation (PBE-GGA)[37]. The plane-wave basis set is converged using a 500 eV energy cutoff. A Γ-centered 3×3×1 *k*-mesh is used for the Brillouin-zone integrations. A Gaussian smearing of 0.02 eV is used for the initial occupations. It is worth pointing out that van der Waals force is excluded from our calculation. We do so not only because strong bonding exits between benzene molecule and $Fe_4N$ (as shown in the due discussion), but also recent studies suggest that it has a negligible effect on GGA optimized structure in, for example, azobenzene/Fe(110)[21].

Bulk $Fe_4N$ has a cubic perovskite-type structure (*Pm3m*) with a lattice constant of 3.795 Å [34]. Fe occupies the corner ($Fe_I$) or face-centered ($Fe_{II}$) position labeled structure graphic of bulk $Fe_4N$, whereas N locates at the body-centered site[34], as shown in Fig. 1(a). Our calculations give a lattice constant of 3.789 Å, in agreement with the experimental value[34]. The $Fe_4N$ substrate is modeled by slabs of three atomic layers with a (3×3) flat surface. We concentrate, in this work, on the effect of different adsorption schemes on the spatial spin-polarization distribution. The subtleties due to the number of atomic layers will be reported in upcoming studies.

## Termination schemes and interface models

The benzene/$Fe_4N$ interfaces are modeled by placing benzene on top of the $Fe_4N$(001) surface. The lattice structure of $Fe_4N$ allows us to have the interfaces with two types of terminations, namely, $Fe_{II}N$ and $Fe_IFe_{II}$. For $Fe_{II}N$ termination, $Fe_4N$ surface is the plane across the body-centered site parallel to $Fe_4N$(001). For $Fe_IFe_{II}$ termination, $Fe_4N$ surface refers to the plane across the face-centered site. For each termination scheme, we further consider two stacking models based on whether N or Fe atom, in the first layer, is right beneath the center of benzene molecule: models $Fe_{II}N$-C and $Fe_IFe_{II}$-C are named after the terminations with N or Fe atom in the $Fe_4N$ surface, locating right beneath the center of benzene molecule, whereas models $Fe_{II}N$-NC and $Fe_IFe_{II}$-NC are referring to the ones without, see Fig. 1. During the lattice structure relaxation, the atoms in the bottom layer of slab are fixed at their bulk positions, whereas other atoms are fully relaxed until the force is weaker than 0.03 eV/Å. In order to decouple adjacent slabs, a thick vacuum layer of 15 Å is included in the direction perpendicular to the surface. To illustrate the nature of the spin-polarization inversion in the real space, we calculate the spin-polarization distribution by the constant-height spin-polarized scanning tunneling microscopy (SP-STM) simulation[38].

## Results and discussion

To illustrate the system, we also define the surface (inter layer, fixed layer) of $Fe_4N$ slab as I (II, III) layer, and the benzene as M layer. We call the zone above benzene as 'benzene surface', between benzene and $Fe_4N$ as 'interfaces', and atoms of I layer, in $Fe_4N$ slab, as '$Fe_4N$ surface'. The sites where the Fe ion located right below the C atom are defined as the top (_t) sites; the ones where the Fe ion located under the C-C bond are defined as the bridge (_b) sites. At the top or bridge sites, we call the Fe (C) atom by Fe_t (C_t) or Fe_b (C_b). The atom located right under the center of benzene is Fe_c or N_c. Figures 1(c), (d), (e) and (f) show the side and top views of the four optimized stacking models. After the structure relaxation, benzene plane is no longer flat in all models except $Fe_{II}N$-NC. Especially, the hydrogen atoms in $Fe_{II}N$-C and $Fe_IFe_{II}$-NC models lie fairly further away from the slab surface than carbon atoms, agreeing with earlier reports[18, 24]. The C-C bonds become longer than those in the isolated benzene ring.

The $Fe_4N$ slab also experiences the structural changes. In both $Fe_{II}N$-C and $Fe_IFe_{II}$-NC models, the Fe_t and Fe_b atoms move out of the Fe plane of the surface, see Fig. 1(c) and (f). In the $Fe_IFe_{II}$-C model, Fe_c moves up, as shown in Fig. 1(e). In the exception arises from the $Fe_{II}N$-NC model, where the benzene plane is still flat, yet the C-C bonds are equal to that in the isolated benzene. The benzene plane is moving away from the $Fe_4N$ surface, as shown in Fig. 1(d). The data from our calculation is in the Table 1.

The adsorption energy ($E_{abs}$) of different models is labeled in Fig. 1(b). According to the adsorption energy, 4 adsorption models fall into two categories: the endothermic adsorption (two $Fe_{II}N$ terminals models) and exothermic adsorption (two $Fe_IFe_{II}$ terminals models). The $Fe_{II}N$-C model has the maximum adsorption energy (0.74 eV). This implies that, at high temperatures, it is the most easily formed model among the four. On the other hand, $Fe_IFe_{II}$-NC shows an exothermic adsorption with the minimum adsorption energy (-2.14 eV), implying that its stability favours low temperatures.

The moment and charge are listed in Table 2. The charge value is calculated using Bader analysis[39-41]. We note that the $Fe_{II}$ moment in layer II, see the 3rd row in Table 2, is smaller than 2.29 $\mu_B$ in bulk $Fe_4N$, where $\mu_B$ is Bohr magneton. This is due to a stronger yet more localized hybridization between N and $Fe_{II}$ in the second layer[42-44]. Apart from the exception in the $Fe_{II}$ ions in layer II, in $Fe_IFe_{II}$ terminal, we observe that, while it gains more charge, the $Fe_{II}$ moment tends to be larger than that in bulk $Fe_4N$. But in the $Fe_{II}N$ terminal, this relation no longer holds; no prominent relationship between charge and moment is present.

To understand the bonding mechanisms, we analyze the charge density difference defined by $\Delta\rho = \rho_{C_6C_6/Fe_4N} - \rho_{C_6H_6} - \rho_{Fe_4N}$, where $\rho_{C_6C_6/Fe_4N}$, $\rho_{C_6H_6}$ and $\rho_{Fe_4N}$ are the charge densities of the full system, isolated benzene and $Fe_4N$ surface, respectively. Charge accumulation (depletion) is in yellow (blue). In the $Fe_{II}N$-C model, the charge accumulates on the C-Fe bonds, as Fig. 2(a) shows. In Fig. 2(b), the interface has little charge accumulation between C and Fe ions, indicating that C atoms do not form bonds with $Fe_4N$ slab. This is consistent with the large distance between the benzene and $Fe_4N$ surface. Figure 2(c) displays a large charge accumulation in the region right below benzene in the $Fe_IFe_{II}$-C model. In the $Fe_IFe_{II}$-NC model, the charge depletion distributes around the centerline perpendicular to $Fe_4N$ surface, and significant charge accumulation appears around the C_t-Fe_t and C_b-Fe_b bonds, as shown in Fig. 2(d). The C atom captures only approximately 0.1~0.2|e|, suggesting the covalent characteristics of the C-Fe bonds.

A general picture of bonding mechanism between the benzene and different $Fe_4N$ terminations can be further extracted from the spin-resolved density-of-states (DOS), as Figs. 3 and 4 show. The 3*d* orbitals can be divided into two classes according to the symmetry: the out-of-plane orbitals ($d_{z^2}$, $d_{xz}+d_{yz}$) and the in-plane ones ($d_{xy}+d_{x^2-y^2}$). In the $Fe_{II}N$-C model, for the top sites, Fe_t $d_{z^2}$ and $d_{xz}+d_{yz}$ hybridize with C_t $p_z$ in the energy interval of -4.25~-3.5 eV for both the spin-up and spin-down states. At about 2.79 eV, the hybridization is just for the spin down states. Meanwhile, the Fe_t spin-up $d_{z^2}$ has hybridization with C_t $p_z$ at 1.80 eV.

For the bridge sites, we note that the spin-up and spin-down π orbitals of benzene are mixed with Fe_b $d_{z^2}$ and $d_{xz}+d_{yz}$ in -4.14~-3.5 eV energy interval, and with Fe_b spin-down $d_{xz}+d_{yz}$ at 2.54 eV. Besides the strong hybridization that mentioned above, a series of hybridizations between the C $p_z$ and Fe *d* states are drawn in Fig. 3(a). For Fe at both the top and bridge sites, its *s* and *d* orbitals, for both spin species, hybridize with the N_c *p* orbitals in the energy interval -7.6~-3.8 eV. The degenerated $p_x+p_y$ orbitals of N_c hybridize with Fe $d_{xy}+d_{x^2-y^2}$ in the energy interval 2.0~5.0 eV. And this hybridization between the spin-down $p_z$ of C and $d_{z^2}$, $d_{xy}+d_{x^2-y^2}$ of Fe_t and Fe_b is fairly strong at -0.43 eV.

In Fig. 3(b) for the $Fe_{II}$N-NC model, the benzene π orbitals originating from the C $p_z$ orbitals do not hybridize with Fe. The slab keeps mostly the properties of a clean surface. The N $p_x$ and $p_y$ orbitals are degenerate. The DOS of Fe_t is almost the same as that of Fe_b. Meanwhile, the $Fe_{II}$ $d_{xz}$+$d_{yz}$ orbitals hybridize with the N $p_z$ orbitals in the energy interval -4.1~-6.6 eV and at Fermi energy ($E_F$). The $Fe_{II}$ $d_{xy}+d_{x^2-y^2}$ and $s$ orbitals are mixed with the degenerate N $p_x$, $p_y$ orbitals at -6.6~-7.2 eV.

In the $Fe_IFe_{II}$ terminations, the intensity of local benzene π orbitals peak become weak gradually, and the peak becomes wider. Figure 4 shows the DOS of two $Fe_IFe_{II}$ terminations. In the $Fe_IFe_{II}$-C model, Fe_c $d_{xz}$+$d_{yz}$ hybridizes with C $p_z$ at -5.1 eV, and with C $p_x$, $p_y$ at -7.9 eV, as shown in Fig. 4(a). The Fe_c $d_{z^2}$ orbitals hybridize with C $p_z$ in the energy interval -6.7~-6.3 eV. The hybridization between the Fe_c spin-up $d_{xy}+d_{x^2-y^2}$ and C $p_z$ is strengthened in the energy interval of 1.1~2.1 eV. At the energy level above 2.5 eV, Fe_c spin-down $d_{z^2}$ weakly hybridizes with C_t $p_z$.

In the $Fe_IFe_{II}$-NC model, we see a rather weak mixture between the C_t $p_z$ and Fe_t $d_{z^2}$, $d_{xz}$+$d_{yz}$ states in the interval -7.4~-6.0 eV. We note that C_t $p_z$ orbitals tend to degenerate with the $p_x$ orbitals, yet C_b shows no such tendency. In the interval -5.5~-4.0 eV, the prime conjugate peaks consist of Fe_b $s$ and C_b $p_z$ orbitals. At -1.75 eV, the Fe_b spin-down $d_{xz}$+$d_{yz}$ and $d_{xy}+d_{x^2-y^2}$ have hybridization with the C_b $p_z$ orbitals.

We note a trend from the above hybridization schemes. As the benzene molecule moves towards the $Fe_4N$ surface, the hybridization of different orbitals depends on the termination schemes. In the $Fe_{II}N$-C and $Fe_IFe_{II}$-C models, the Fe $d_{xz}+d_{yz}$ and $d_{z^2}$ orbitals hybridize strongly with the C $p_z$ state, leading to spin-polarization inversion. In the $Fe_IFe_{II}$-NC model, the hybridization between both spin species of the C $p_z$ and Fe $s$, $d_{xy}+d_{x^2-y^2}$ orbitals is stronger than that between C $p_z$ and Fe ($d_{xz}+d_{yz}$, $d_{z^2}$), unable to reverse the spin polarization. This is consistent with the report in Ref. 17. We are thus led to conclude that the spin-polarization inversion at benzene surface is a result of hybridizations between the $p_z$ orbital of C and the out-of-plane Fe $d$ orbitals.

We show, in Fig. 5, the spatial distribution of spin-polarization $P_{space}$, defined as

$$P_{space} = \frac{n_s^{\uparrow}(\mathrm{r}_{\parallel}, z, \varepsilon) - n_s^{\downarrow}(\mathrm{r}_{\parallel}, z, \varepsilon)}{n_s^{\uparrow}(\mathrm{r}_{\parallel}, z, \varepsilon) + n_s^{\downarrow}(\mathrm{r}_{\parallel}, z, \varepsilon)}, \quad (1)$$

for an energy interval of [$\varepsilon$, $E_F$]. $n_s^{\uparrow\downarrow}(\mathrm{r}_{\parallel}, z, \varepsilon)$ is the spin-up (down) charge density in real space, at position $\mathrm{r}_{\parallel}$ and a distance $z$ from the surface. Here, the value of $\varepsilon$ is either $E_F$-0.4 eV or $E_F$+0.4 eV [21, 28]. In this figure, we focus on two energy intervals, [$E_F$-0.4 eV, $E_F$] and [$E_F$, $E_F$+0.4 eV]. In each energy interval, the spin-polarization is projected onto the plane that is parallel to the $Fe_4N$ surface, see Figs. 5(b) and (c); the distance between the plane and benzene surface is labeled in Fig. 5(b). For the 4 models discussed in this work, we plot, in Figs. 5(a) and (d), the spin polarization along a few selected lines defined in Figs. 5(b) and (c).

For the $Fe_{II}N$-C model, the highest spin polarization is ~80%, and the lowest value is ~-60% via line 2 and line 3, see Fig. 5(a). The intensity of inversion is much stronger than benzene adsorbed antiferromagnetic Mn[26], due to the hybridization between the $p_z$ states and $d_{z^2}$, $d_{xy}+d_{x^2-y^2}$ orbitals of Fe_t and Fe_b at -0.43 eV. This hybridization enhances the population of the spin-down species, and is thus reversing the spin polarization. It is interesting to note that the spin-polarization distributions in the energy interval [$E_F$-0.4 eV, $E_F$] and [$E_F$, $E_F$+0.4 eV] are rather different, even with opposite signs. This suggests that the sign of spin polarization can be reversed by simply shifting the $E_F$ by, for example, applying a gate voltage.

In the $Fe_{II}N$-NC model, line 2 exhibits the spin-polarization inversion, as shown in Fig. 5(b). The DOS, in Fig. 3(b), however, points to a weak adsorption. The spin-polarization distribution in this model is thus similar to that in vacuum (above a clean $Fe_4N$ surface). For the $Fe_{II}N$-NC model, the spin polarization of line 1 with a cosine-type distribution is shown in Fig. 5(a).

In both $Fe_IFe_{II}$ terminal models, spin-polarization inversion happens, but the strong spin-polarization inversion in the neighbourhood of benzene happens only in $Fe_IFe_{II}$-C. In $Fe_IFe_{II}$ terminations, $Fe_4N$ surface distort significantly, $Fe_{II}$ ions are not located right above N atoms, as shown in Figs. 1(e) and (f). Then positive spin polarization of N atoms extends into benzene surface. In $Fe_IFe_{II}$-C model, the most interesting feature is that the positive spin-polarization distributes along the C-C bonds, see Fig. 5(b). In Fig. 6, a positive spin polarization of benzene appears at $E_F$ in the $Fe_IFe_{II}$-C model. On the other hand, in the $Fe_IFe_{II}$-NC model spin polarization is approximate 0%. Meanwhile spatial spin-polarization, for $Fe_IFe_{II}$-NC, in the neighbourhood of benzene is almost 0%, see Fig. 5(b). So, the atomic scale spin-polarization, at benzene surface, is modulated by N and C atoms.

Figure 7 is the spin-polarization plane of $Fe_{II}$N-C structure. It's across the top sites and parallel to $Fe_4N$(100). From this figure, benzene hampers the extension of N position spin polarization, and realizes the spatial spin polarization inversion. The reason is the overlap of $p_z$ and out-of-plane components of $d$.

## Conclusion

In summary, we have shown that at the spinterface formed by benzene adsorbed on $Fe_4N$, depending on the specific termination schemes, a variety of spin polarization, including spin polarization inversion, can take place. The spin-polarization inversion finds its origin in the hybridization between the out-of-plane components of Fe $d$ orbitals and the benzene $\pi$ orbitals (the $p_z$ orbital, in particular). The presence of N atoms partition the adsorption into two categories: the endothermic (adsorption) $Fe_{II}$N terminal models and the exothermic $Fe_IFe_{II}$ terminal ones. With these results, we can see that adsorptions rely on the temperature. The $Fe_{II}$N-C with the maximum adsorption energy will be easier to be formed than others under high temperature and has significant spin-polarization inversion, which is desired for the spintronic devices.

## Acknowledgements

This work is supported by the National Natural Science Foundation of China (51171126), Key Project of Natural Science Foundation of Tianjin City (12JCZDJC27100 and 14JCZDJC37800), Program for New Century Excellent Talents in University (NCET-13-0409), Scientific Research Foundation for the Returned Overseas Chinese Scholars, State Education Ministry. It is also supported by High Performance Computing Center of Tianjin University, China.

## Author contributions

All authors designed the outline of the manuscript. Q.Z. and W.B.M. wrote the main text; X.C.W. and X.H.W. contributed detailed discussions and revisions; All authors reviewed the manuscript.

# Competing financial interests statement

Competing financial interests:

The authors declare no competing financial interests.

# Table notes

**Table 1.** Special bonds of different adsorbed structures comparing to the clear $Fe_4N$ surface. The bonds of isolated benzene is 1.398 Å. X-up-Y means that the average vertical distance between X and Y plane. When Y=Fe, Y stands for the Fe atoms in layer I.

| | Adsorbed | | | | Clear | |
|---|---|---|---|---|---|---|
| Bond (Å) | $Fe_{II}N$-C | $Fe_{II}N$-NC | $Fe_IFe_{II}$-C | $Fe_IFe_{II}$-NC | $Fe_{II}N$ | $Fe_IFe_{II}$ |
| IL-IIL | 1.674 | 1.636 | 1.772 | 1.834 | 1.701 | 1.706 |
| IIL-IIIL | 1.836 | 1.813 | 1.897 | 1.872 | 1.754 | 1.844 |
| N-up-Fe | 0.292 | 0.344 | - | - | 0.343 | (Fe)0.200 |
| M-IL | 2.353 | 3.634 | 1.997 | 1.983 | - | - |
| C_t-C_b | 1.442 | 1.398 | 1.418 | 1.442 | - | - |
| C_b-C_b | 1.408 | 1.399 | 1.437 | 1.463 | - | - |
| H-up-C | 0.266 | 0 | 0.098 | 0.360 | - | - |
| Fe_t-up-Fe | 0.319 | 0 | - | 0.139 | - | - |
| Fe_b-up-Fe | 0.253 | 0 | - | -0.05 | - | - |
| Center-up-Fe | (N)0.140 | - | (Fe)0.317 | - | - | - |

**Table 2.** The average moment (Mom $\mu_B$) and charge (Chg e) of four models. The charge of $Fe_{II}$ in bulk $Fe_4N$ is 7.61 |e|.

| | $Fe_{II}N$-C | | $Fe_{II}N$-NC | | $Fe_IFe_{II}$-C | | $Fe_IFe_{II}$-NC | |
|---|---|---|---|---|---|---|---|---|
| | Mom ($\mu_B$) | Chg (e) | Mom ($\mu_B$) | Chg (e) | Mom ($\mu_B$) | Chg (e) | Mom ($\mu_B$) | Chg (e) |
| I-$Fe_{II}$ | 2.302 | 7.466 | 2.294 | 7.463 | 2.497 | 7.756 | 2.540 | 7.729 |
| I-N($Fe_I$) | -0.040 | 6.227 | -0.048 | 6.211 | 2.905 | 7.896 | 2.835 | 7.896 |
| II-$Fe_{II}$ | 1.514 | 7.684 | 1.268 | 7.730 | 2.049 | 7.656 | 2.078 | 7.660 |
| II-N($Fe_I$) | 2.844 | 7.880 | 2.823 | 7.893 | -0.016 | 6.246 | -0.016 | 6.243 |
| III-$Fe_{II}$ | 2.360 | 7.474 | 2.330 | 7.513 | 2.731 | 7.827 | 2.714 | 7.841 |
| III-N($Fe_I$) | -0.039 | 6.293 | -0.043 | 6.214 | 3.011 | 7.905 | 3.006 | 7.890 |
| C_t | -0.034 | 4.150 | 0.002 | 4.055 | 0.008 | 4.128 | -0.018 | 4.199 |
| C_b | -0.010 | 4.127 | 0.002 | 4.071 | -0.009 | 4.188 | -0.017 | 4.229 |
| H | 0.001 | 0.920 | 0.000 | 0.934 | 0.003 | 0.918 | 0.003 | 0.901 |
| Center | -0.061 | 6.228 | - | - | 1.955 | 7.484 | - | - |
| Fe_b | 2.250 | 7.404 | 2.298 | 7.452 | 2.048 | 7.637 | 2.786 | 7.746 |
| Fe_t | 2.296 | 7.368 | 2.298 | 7.453 | 2.073 | 7.639 | 2.372 | 7.537 |

## Figure captions

**Figure 1.** The structure of bulk $Fe_4N$ and the side and top views of four benzene/$Fe_4N$(001) structures. The adsorption energies ($E_{abs}$) are labeled in Fig. 1(b).

**Figure 2.** The side and top views of charge density difference in four benzene/$Fe_4N$(001) structures. Yellow (blue) regions represent the net charge gain (loss).

**Figure 3.** The spin resolved density of states of the $Fe_{II}$N-C and $Fe_{II}$N-NC models.

**Figure 4.** The spin resolved density of states of the $Fe_IFe_{II}$-C and $Fe_IFe_{II}$-NC models.

**Figure 5.** Spin polarization nearby the benzene plane in the vacuum for these four models. (a) and (b) represent spin polarization distribution in [$E_F$-0.4 eV, $E_F$] energy interval; (c) and (d) represent spin-polarization distribution in [$E_F$, $E_F$+0.4 eV] energy interval. (a)/(d) is line profiles of the spin-polarization selected in (b)/(c), for different structures, respectively. The height of the spin-polarization distribution plane relative to the benzene plane are labeled

under (b) for each model.

**Figure 6.** The spin-resolved density of states of benzene adsorbed on $Fe_IFe_{II}$ surface in the $E_F$-1~$E_F$+1 eV interval.

**Figure 7.** The spin-polarization distribution vertical to $Fe_4N$ surface for the $Fe_{II}$N-C structure. This plane is parallel to $Fe_4N(100)$ and across the top site. The C_t atom is labeled and the value of $\varepsilon$ is $E_F$-0.4 eV.

**Figure 1, Q. Zhang *et al.***

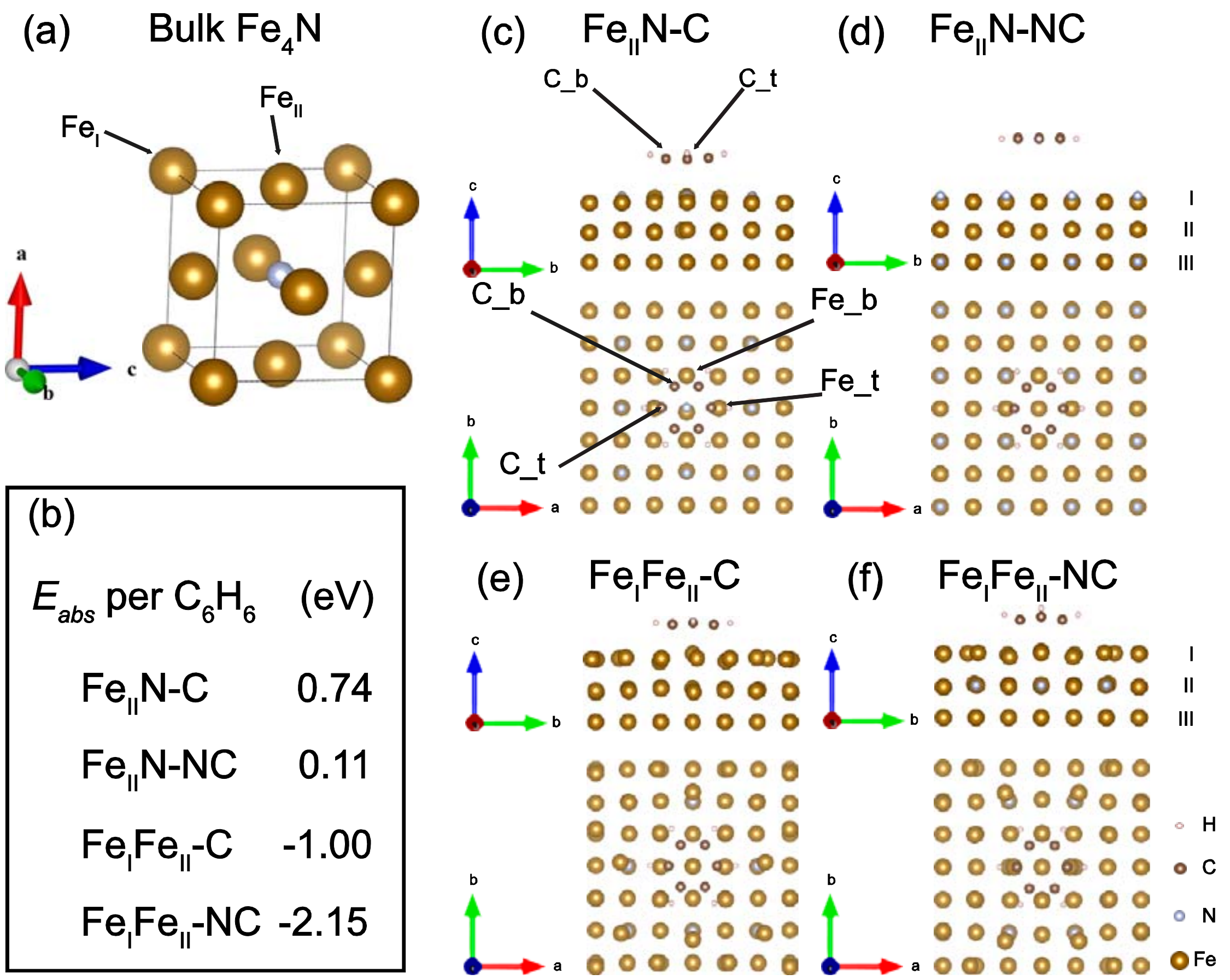

**Figure 2, Q. Zhang *et al.***

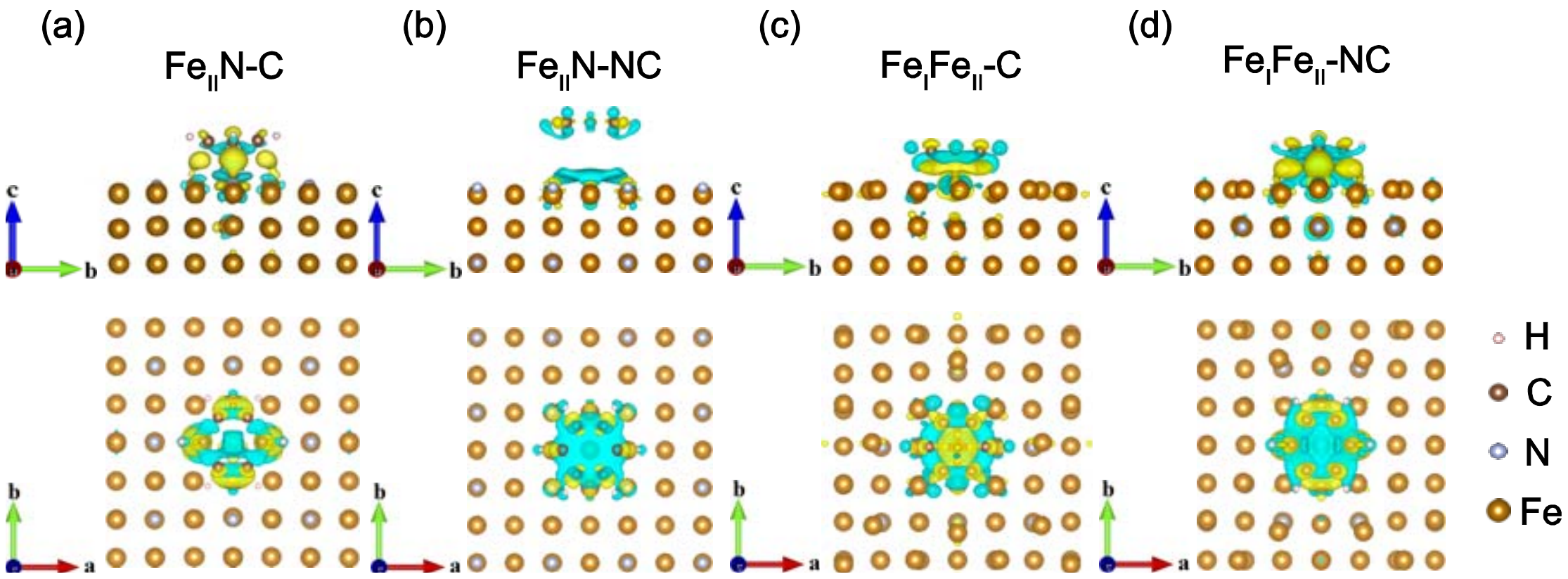

**Figure 3, Q. Zhang *et al.***

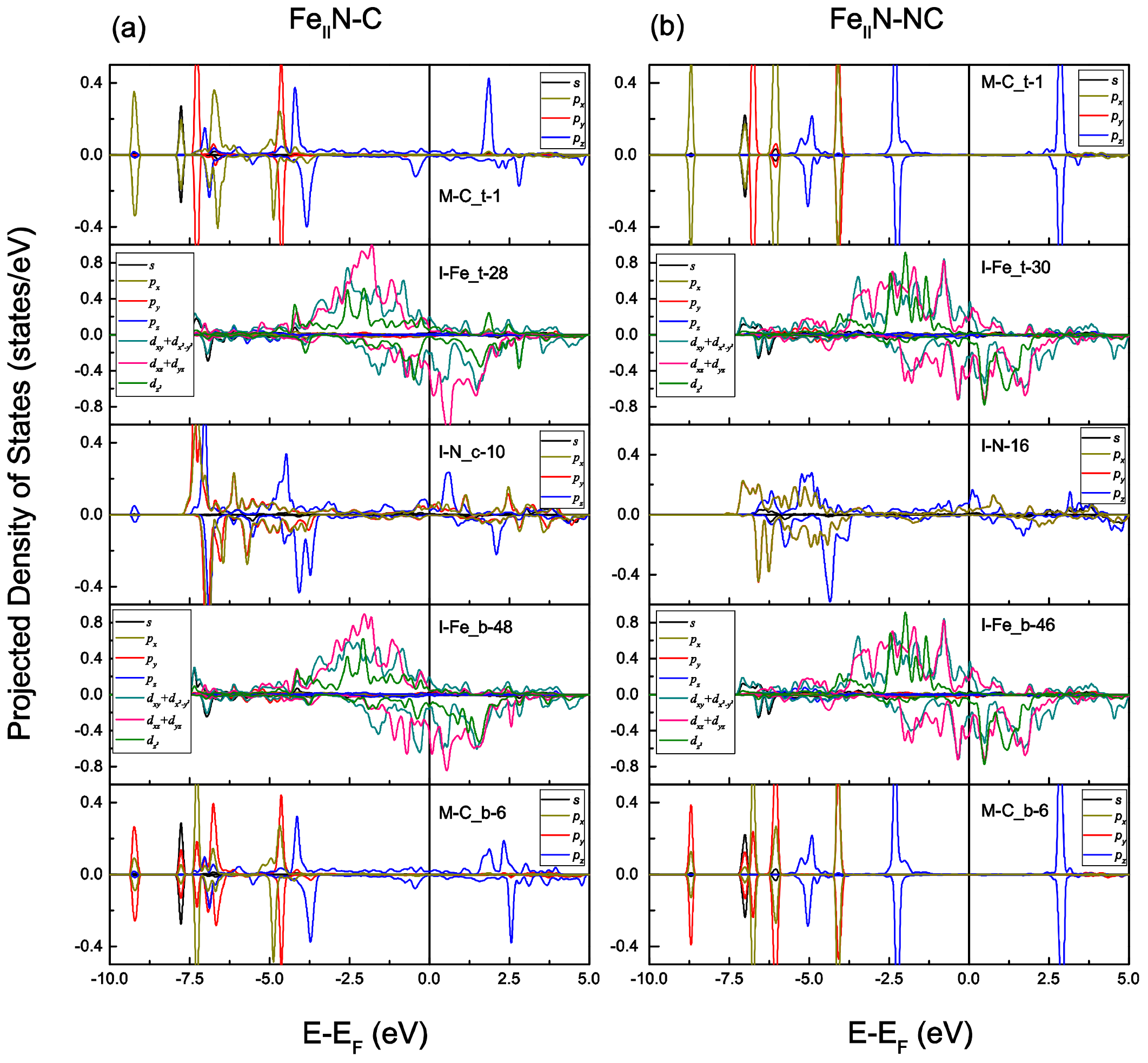

**Figure 4, Q. Zhang *et al.***

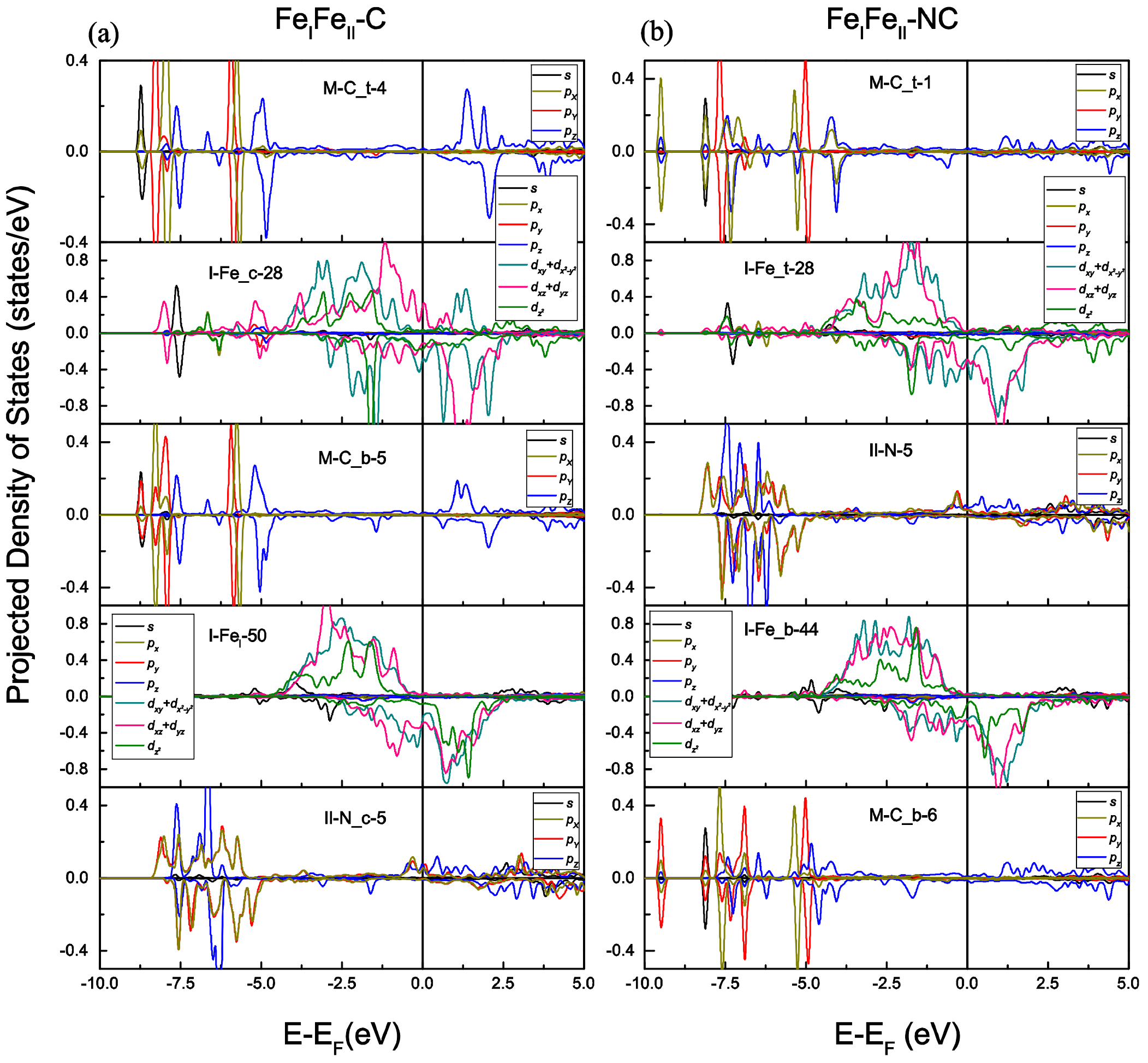

**Figure 5, Q. Zhang *et al.***

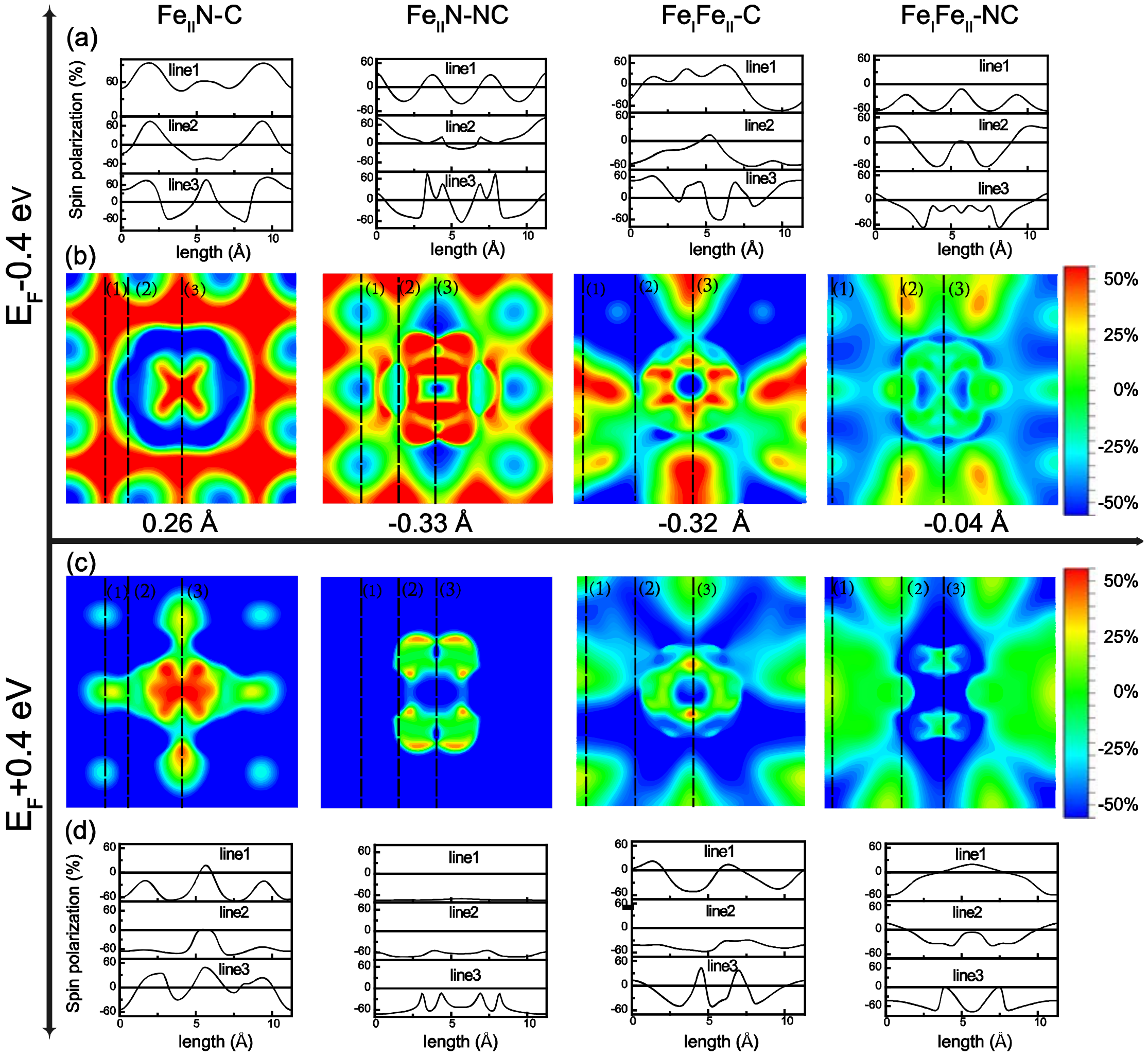

**Figure 6, Q. Zhang *et al.***

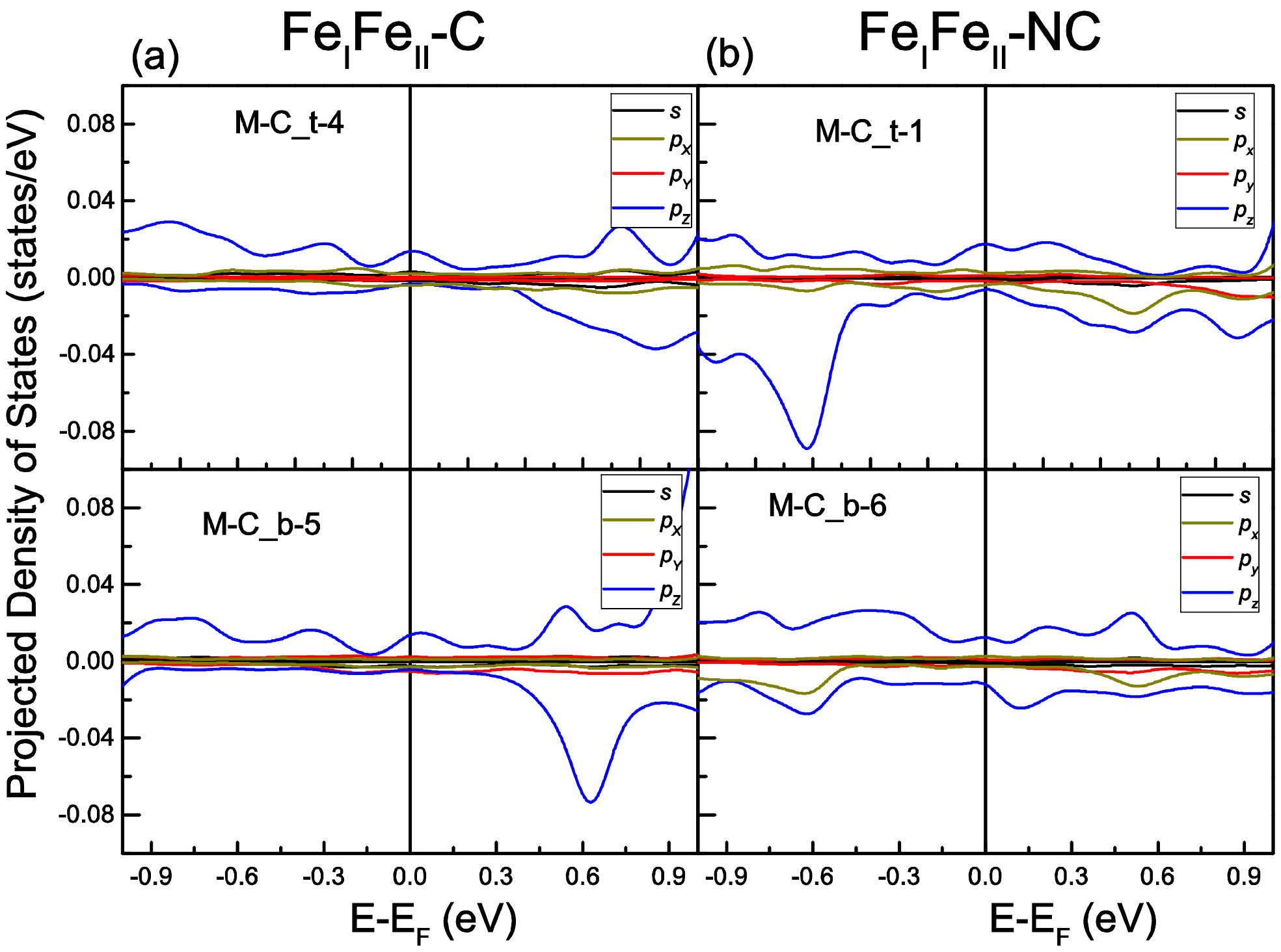

**Figure 7, Q. Zhang *et al.***

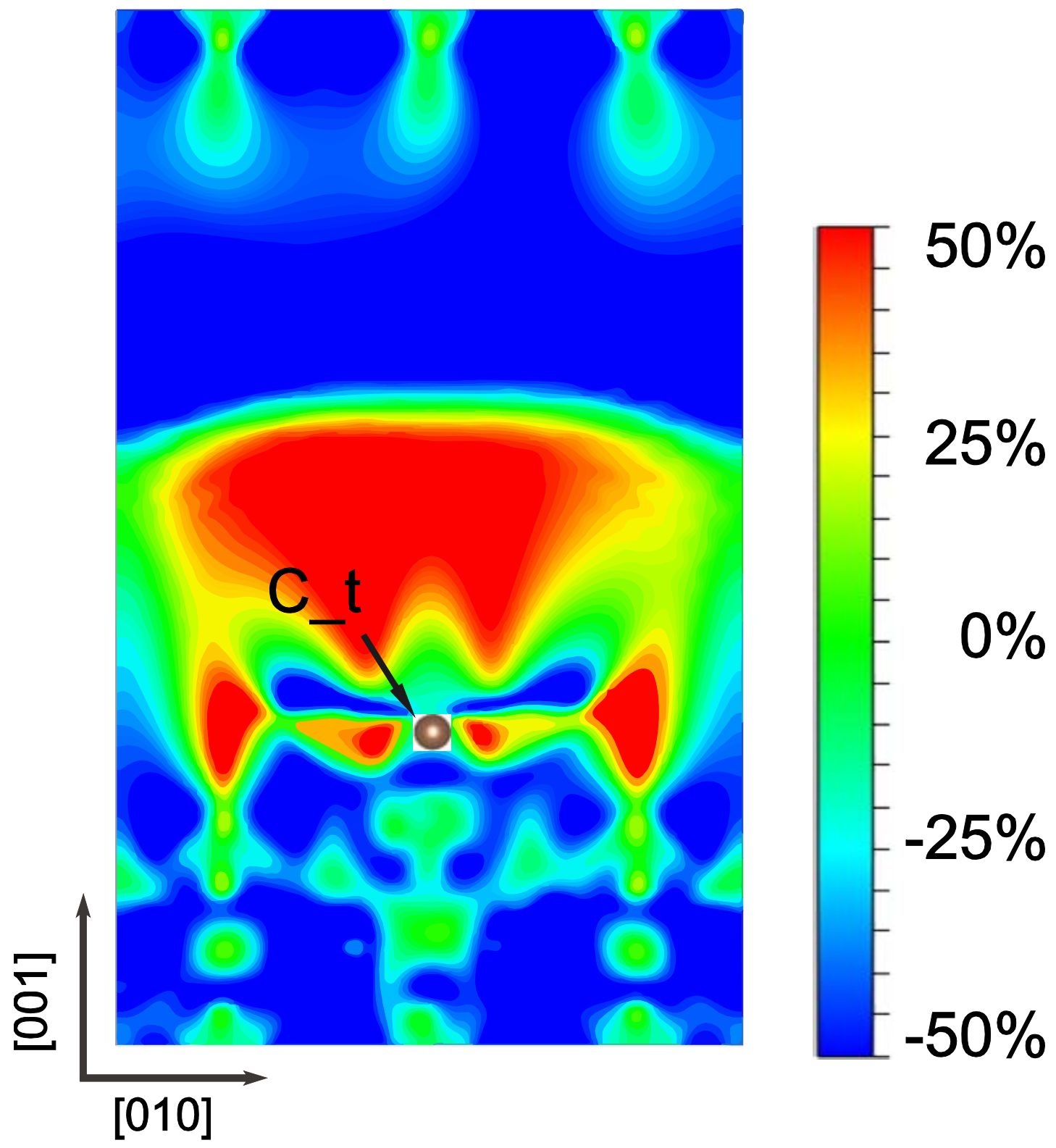